\documentclass[10pt]{article}
\usepackage[french,english]{babel}
\usepackage{amsmath}
\usepackage{amsfonts}
\usepackage{amssymb}
\usepackage{times}
\usepackage{version}
\begin{document}
%
%%%%%%%%%%%%%%%%%%%%%%%%%%%%%%%%%%%%%%%%%%%%%%%%%%%%%%%%%%%%%%%%%%%%%%%%%%%%%%
%
January 9th, 2013   \hfill Cab2gen3.tex

\hfill arXiv:1212.0627

%\today\quad   \hfill Cab2gen3.tex
\vskip 4.2cm
{\baselineskip 12pt
\begin{center}
{\bf UNLOCKING THE STANDARD MODEL}
\end{center}
\begin{center}
        {\bf III .\quad 2 GENERATIONS OF QUARKS : CALCULATING THE CABIBBO ANGLE}
\end{center}
}
\baselineskip 16pt
%=====================================================================
\arraycolsep 3pt  % space between columns in matrices
%=====================================================================
%
\vskip .2cm
\centerline{B.~Machet
     \footnote[1]{LPTHE tour 13-14, 4\raise 3pt \hbox{\tiny \`eme} \'etage,
          UPMC Univ Paris 06, BP 126, 4 place Jussieu,
          F-75252 Paris Cedex 05 (France),\\
         Unit\'e Mixte de Recherche UMR 7589 (CNRS / UPMC Univ Paris 06)}
    \footnote[2]{machet@lpthe.jussieu.fr}
     }
\vskip 1cm

{\bf Abstract:} Maximally extending the Higgs sector of the
Glashow-Salam-Weinberg model by including all scalar and pseudoscalar $J=0$
states expected for 2 generations of quarks,
I demonstrate that the Cabibbo angle is given by
$\tan^2\theta_c=
\displaystyle\frac{\frac{1}{m_K^2}-\frac{1}{m_D^2}}
{\frac{1}{m_\pi^2}-\frac{1}{m_{D_s}^2}}
\approx\frac{m_\pi^2}{m_K^2}\left(1-\frac{m_K^2}{m_D^2}
+\frac{m_\pi^2}{m_{D_s}^2}\right)$.

\smallskip


%%%%%%%%%%%%%%%%%%%%%%%%%%%%%%%%%%%%%%%%%%%%%%%%%%%%%%%%%%%%%%%%%%%%%%%%%%%%%%%
%%%%%%%%%%%%%%%%%%%%%%%%%%%%%%%%%%%%%%%%%%%%%%%%%%%%%%%%%%%%%%%%%%%%%%%%%%%%%%%
\section{Introduction} \label{section:intro}
%===========================================

In \cite{Machet1} and \cite{Machet2}, I proposed to minimally extend the
Glashow-Salam-Weinberg (GSW) model \cite{GSW} by maximally enlarging its Higgs sector,
including  in there all $J=0$ scalar (and pseudoscalar) states that
can be expected for a given number $N$ of generation of quarks.
The $8N^2$ such states, transforming like $\bar q_i q_j$ or $\bar q_i
\gamma^5 q_j$ composite operators and suitably normalized can be divided
into $2N^2$ quadruplets which are all in one-to-one correspondence with the
complex Higgs doublet of the GSW model \cite{Machet}
\footnote{Some normalization factors are  erroneous in \cite{Machet}
but have been corrected here.}.
The works \cite{Machet1} and \cite{Machet2} were dedicated to the
restrictive  case of 1 generation. Here I focus on the 2-generations case,
but only present the  calculation of the Cabibbo angle, 
leaving a more detailed exposition  to a longer work \cite{Machet4}.

\section{Laws of transformation and isomorphism} \label{section:laws}
%====================================================================

\subsection{The complex Higgs doublet of the Glashow-Salam-Weinberg model}
%------------------------------------------------------------------------

If, instead of the customary form $H=\left(\begin{array}{c}
\chi^1+i\chi^2 \cr \chi^0 -ik^3 \end{array}\right)$ involving the 4 reals
fields $\chi^0, \chi^1,\chi^2, k^3=-i\chi^3$,
the complex scalar doublet $H$ of the GSW model is written
\begin{equation}
H=\left(\begin{array}{r}
h^1-ih^2\cr -(h^0+h^3)\end{array}\right),
\label{eq:doublet}
\end{equation}
the laws of transformation of its $h^0, h^j, j=1,2,3$ components by the group
 $SU(2)_L$ with generators $T^i_L,\ i=1,2,3$ write
\footnote{A transformation ${\cal U}_L$ of the $SU(2)_L$ group is written
${\cal U}_L = e^{-i\alpha_i T^i_L}, i=1,2,3$.}
\begin{equation}
\boxed{
\begin{array}{rcl}
T^i_L\,.\,h^j&=&-\frac{1}{2}\left(i\,\epsilon_{ijk}h^k +
\delta_{ij}\,h^0\right)\cr
T^i_L\,.\,h^0 &=& -\frac{1}{2}\, h^i
\end{array}
}
\label{eq:ruleL}
\end{equation}
Acting in the space of quark flavors $(u,c,d,s)$ with dimension $2N=4$, the three
$SU(2)$ generators can be represented by 
\begin{equation}
T^3 = \frac 12 \left(\begin{array}{ccc} {\mathbb I} & \vline & \\
\hline
                      & \vline & -{\mathbb I}\end{array}\right),
\quad
T^+ = T^1+iT^2= \left(\begin{array}{ccc}  & \vline &{\mathbb I} \\
\hline
                      & \vline & \end{array}\right),
\quad
T^- = T^1-iT^2= \left(\begin{array}{ccc}  & \vline & \\
\hline
                      {\mathbb I}& \vline & \end{array} \right),
\end{equation}
where $\mathbb I$ is the $N\times N = 2\times 2$ identity matrix.
So doing, we realize an embedding of $SU(2)_L$ and/or $SU(2)_R$ into the
chiral group $U(2N)_L \times U(2N)_R$.

\subsection{Composite Higgs doublets}
%-----------------------------------

We  now act with this chiral group on composite operators of the form
$\bar\psi {\mathbb M}\psi$ and $\bar \psi \gamma^5 {\mathbb M}\psi$, where
$\psi$ is the $2N$-vector of flavor quarks $\psi = (u,c,d,s)^T$ and
$\mathbb M$ is any $2N\times 2N (=4\times 4)$ matrix. 
\begin{eqnarray}
({\cal U}_L \times {\cal U}_R)\,.\,\bar\psi\frac{1+\gamma_5}{2}{\mathbb M}\psi
&=& \bar \psi\; {\cal U}_L^{-1}\,{\mathbb M\;{\cal U}_R\;\frac{1+\gamma_5}{2}}\psi,\cr
&& \cr
({\cal U}_L \times {\cal U}_R)\,.\,\bar\psi\frac{1-\gamma_5}{2}{\mathbb M}\psi
&=& \bar \psi\; {\cal U}_R^{-1}\,{\mathbb M\;{\cal U}_L\;\frac{1-\gamma_5}{2}}\psi.
\label{eq:group}
\end{eqnarray}
Writing left and right transformations of the group as
\begin{equation}
{\cal U}_{L,R} = e^{-i \alpha_i T^i_{L,R}},\quad i=1,2,3
\end{equation}
eq.~(\ref{eq:group}) entails ($[\ ,\ ]$ and $\{\ ,\ \}$ stand respectively
for the commutator and anticommutator)

\vbox{
\begin{eqnarray}
{ T}^j_L\,.\,\bar\psi{\mathbb M} \psi &=&
-\frac12\,\left(\bar\psi\,[{ T}^j,{\mathbb M}]\,\psi
                 +\bar\psi\, \{{ T}^j,{\mathbb M}\}\,\gamma_5\psi\right),\cr
&& \cr
{ T}^j_L\,.\,\bar\psi{\mathbb M}\gamma_5 \psi &=&
-\frac12\,\left(\bar\psi\,[{ T}^j,{\mathbb
M}]\,\gamma_5\psi
                 +\bar\psi\, \{{ T}^j,{\mathbb M}\}\,\psi\right),\cr
&& \cr
{ T}^j_R\,.\,\bar\psi{\mathbb M} \psi &=&
-\frac12\,\left(\bar\psi\,[{ T}^j,{\mathbb M}]\,\psi
                 -\bar\psi\, \{{ T}^j,{\mathbb M}\}\,\gamma_5\psi\right),\cr
&& \cr
{ T}^j_R\,.\,\bar\psi{\mathbb M}\gamma_5 \psi &=&
-\frac12\,\left(\bar\psi\,[{ T}^j,{\mathbb
M}]\,\gamma_5\psi
                 -\bar\psi\, \{{ T}^j,{\mathbb M}\}\,\psi\right).
\label{eq:trans2}
\end{eqnarray}
}

Let us consider the following set of $2N^2=8$ quadruplets (${\mathbb
M}^\pm={\mathbb M}^{1\pm i2}$)
\begin{equation}
\bar\psi \Big({\mathbb M}^0, \gamma^5{\mathbb M}^3, \gamma^5{\mathbb
M}^+, \gamma^5{\mathbb M}^-\Big)\psi
\label{eq:SP}
\end{equation}
and
\begin{equation}
\bar\psi \Big(\gamma^5{\mathbb M}^0, {\mathbb M}^3, {\mathbb M}^+,
{\mathbb M}^-\Big)\psi,
\label{eq:PS}
\end{equation}
in which
\begin{equation}
{\mathbb M}^0= \left(
\begin{array}{ccc} M & \vline & 0 \cr
\hline
0 & \vline & M \end{array}\right),
{\mathbb M}^3 = \left(
\begin{array}{ccc} M & \vline & 0 \cr
\hline
0 & \vline & -M \end{array}\right),
 {\mathbb M}^+ =2 \left(
\begin{array}{ccc} 0 & \vline & M \cr
\hline
0 & \vline & 0 \end{array}\right),
{\mathbb M}^-= 2\left(
\begin{array}{ccc} 0 & \vline & 0 \cr
\hline
M & \vline & 0 \end{array}\right),
\end{equation}
$M$ being any $N\times N=2\times 2$ real matrix.
Denoting generically these quadruplets $A$ and their  components 
$(a^0, a^3, a^+, a^-)$, their laws of transformations by
$SU(2)_L$ are given by (\ref{eq:ruleL}), in which  $h^0, h^i$ has been
replaced by $a^0, a^i$, while they transform by $SU(2)_R$ according to
\begin{equation}
\begin{array}{rcl}
T^i_R\,.\,a^j&=&-\frac{1}{2}\left(i\,\epsilon_{ijk}a^k -
\delta_{ij}\,a^0\right),\cr
T^i_R\,.\,a^0 &=& +\frac{1}{2}\, a^i.
\end{array}
\label{eq:ruleR}
\end{equation}
We have therefore found $2N^2$ ``composite'' quadruplets isomorphic to the complex
doublet of the GSW model. They split into $N^2$ of the type $({\mathfrak
s}^0, \vec {\mathfrak p})$
and $N^2$ of the type $({\mathfrak p}^0, \vec {\mathfrak s})$, in which ${\mathfrak s}$ stands for ``scalar'' and
${\mathfrak p}$ for ``pseudoscalar. These two subsets are transformed into each other
by parity (the corresponding generator being ${\mathbb I}_L$ or ${\mathbb
I}_R)$.
 Their $8N^2$ components span the whole set of
scalar and pseudoscalar $J=0$ composite states that can be ``built'' with
$2N$ quarks. In this sense, this extension represents the maximal possible
extension of the Higgs sector of the GSW model.

\subsection{Normalization}
%-------------------------

All composite operators that have been defined above having dimension
$[mass]^3$, the quadruplets need to be suitably normalized.
To this purpose we introduce $2 \times 2N^2$ parameters corresponding
to the vacuum expectations values (VEV's) of, respectively:\newline
* the scalar neutral composite operator of dimension $[mass]^3$ occurring in
each quadruplet, which can only be ${\mathfrak s}^0$ or ${\mathfrak s}^3$;
this VEV we call $\mu^3$ in the first case and  $\hat\mu^3$ in the second
case, with an index that labels the quadruplet under concern;\newline
* the corresponding scalar ``Higgs'' field with dimension $[mass]$; we call
it $\frac{v}{\sqrt{2}}$ for an ${\mathfrak s}^0$ and $\frac{\hat
v}{\sqrt{2}}$ for an ${\mathfrak s}^3$, with an index $X, H, \Omega$ or
$\Xi$ labeling the quadruplet under concern.              

We thus consider hereafter the $N^2=4$ following  $({\mathfrak s}^0,\vec
{\mathfrak p})$ quadruplets
\begin{equation}
\begin{split}
 X &=\frac{v_X}{\sqrt{2}\,\mu_X^3}\frac{1}{\sqrt{2}}\ \bar\psi\left(
\left(\begin{array}{rrrrr}
1 &  & \vline &  & \cr
 & 0 & \vline & & \cr
\hline
 & & \vline & 1 &  \cr
 & & \vline &  & 0 \end{array}\right),
\gamma^5\left(\begin{array}{rrrrr}
1 &  & \vline &  & \cr
 & 0 & \vline & & \cr
\hline
 & & \vline & -1 &  \cr
 & & \vline &  & 0 \end{array}\right),
2\gamma^5\left(\begin{array}{rrrrr}
 &  & \vline & 1 &  \cr
 &  & \vline &  & 0\cr
\hline
 & & \vline &  &  \cr
 & & \vline &  & \end{array}\right),
2\gamma^5\left(\begin{array}{rrrrr}
 &  & \vline &  &  \cr
 &  & \vline &  & \cr
\hline
1 &   & \vline &  &  \cr
 & 0  & \vline &  & \end{array}\right)
\right)\psi\cr
&= (X^0,X^3,X^+,X^-),
 \quad with\quad \mu_X^3 = \frac{<\bar u u+\bar d d>}{\sqrt{2}},
\end{split}
\label{eq:Xq}
\end{equation}

\begin{equation}
\begin{split}
 H &=\frac{v_H}{\sqrt{2}\,\mu_H^3}\frac{1}{\sqrt{2}}\ \bar\psi\left(
\left(\begin{array}{ccccc}
0 &  & \vline &  & \cr
 & 1 & \vline & & \cr
\hline
 & & \vline & 0 &  \cr
 & & \vline &  & 1 \end{array}\right),
\gamma^5 \left(\begin{array}{ccccc}
0 &  & \vline &  & \cr
 & 1 & \vline & & \cr
\hline
 & & \vline & 0 &  \cr
 & & \vline &  & -1 \end{array}\right),
2\gamma^5 \left(\begin{array}{ccccc}
 &  & \vline & 0 &  \cr
 &  & \vline &  & 1\cr
\hline
 & & \vline &  &  \cr
 & & \vline &  & \end{array}\right),
2\gamma^5 \left(\begin{array}{ccccc}
 &  & \vline &  &  \cr
 &  & \vline &  & \cr
\hline
0 &   & \vline &  &  \cr
 & 1  & \vline &  & \end{array}\right)
\right)\psi\cr
&= (H^0,H^3,H^+,H^-),
 \quad with\quad \mu_H^3 = \frac{<\bar c c+\bar s s>}{\sqrt{2}},
\end{split}
\label{eq:Hq}
\end{equation}

\begin{equation}
\begin{split}
 \Omega
&=\frac{v_\Omega}{\sqrt{2}\,\mu_\Omega^3}\frac12\ \bar\psi\left(
\left(\begin{array}{ccccc}
 & 1 & \vline &  & \cr
1 &  & \vline & & \cr
\hline
 & & \vline &  & 1 \cr
 & & \vline & 1 &  \end{array}\right),
\gamma^5\left(\begin{array}{ccccc}
 & 1 & \vline &  & \cr
1 &  & \vline & & \cr
\hline
 & & \vline &  & -1 \cr
 & & \vline & -1 &  \end{array}\right),
2\gamma^5\left(\begin{array}{ccccc}
 &  & \vline &  & 1 \cr
 &  & \vline & 1 & \cr
\hline
 & & \vline &  &  \cr
 & & \vline &  & \end{array}\right),
2\gamma^5\left(\begin{array}{ccccc}
 &  & \vline &  &  \cr
 &  & \vline &  & \cr
\hline
 & 1  & \vline &  &  \cr
1 &   & \vline &  & \end{array}\right)
\right)\psi\cr
&= (\Omega^0,\Omega^3,\Omega^+,\Omega^-),
 \quad with\quad \mu_\Omega^3 = \frac{<\bar u c+\bar c u +\bar d
s+\bar s d>}{2},
\end{split}
\label{eq:Omq}
\end{equation}

\begin{equation}
\begin{split}
\Xi &=\frac{v_\Xi}{\sqrt{2}\,\mu_\Xi^3}\frac12\ \bar\psi\left(
\left(\begin{array}{rrrrr}
 & 1 & \vline &  & \cr
-1 &  & \vline & & \cr
\hline
 & & \vline &  & 1 \cr
 & & \vline & -1 &  \end{array}\right),
\gamma^5\left(\begin{array}{rrrrr}
 & 1 & \vline &  & \cr
-1 &  & \vline & & \cr
\hline
 & & \vline &  & -1 \cr
 & & \vline & 1 &  \end{array}\right),
2\gamma^5\left(\begin{array}{rrrrr}
 &  & \vline &  & 1 \cr
 &  & \vline & -1 & \cr
\hline
 & & \vline &  &  \cr
 & & \vline &  & \end{array}\right),
2\gamma^5\left(\begin{array}{rrrrr}
 &  & \vline &  &  \cr
 &  & \vline &  & \cr
\hline
 & 1  & \vline &  &  \cr
-1 &   & \vline &  & \end{array}\right)
\right)\psi\cr
&= (\Xi^0,\Xi^3,\Xi^+,\Xi^-),
 \quad with\quad \mu_\Omega^3 = \frac{<\bar u c -\bar c u + \bar d s  -\bar
 s d>}{2},
\end{split}
\label{eq:Xiq}
\end{equation}

and their $N^2$ parity transformed $({\mathfrak p}^0, \vec {\mathfrak s})$
quadruplets that we call
$\hat X, \hat H, \hat\Omega, \hat\Xi$. The latter are associated with the VEV's
$\hat v_X, \hat v_H, \hat v_\Omega, \hat v_\Xi$, and
\begin{equation}
\hat\mu_X^3 =\frac{<\bar u u-\bar d d>}{\sqrt{2}},
\hat\mu_H^3=\frac{<\bar c c-\bar s s>}{\sqrt{2}},
\hat\mu_\Omega^3 =\frac{<\bar u c+\bar c u - \bar d s-\bar s d>}{2},
\hat\mu_\Xi^3=\frac{<\bar u c-\bar c u-\bar d s+\bar s d>}{2}.
\end{equation}
We suppose that the VEV's of pseudoscalar neutral
composite operators vanish, which is certainly true at the classical level
(they may receive non-vanishing quantum corrections in a parity
violating theory like this one, but this is beyond the scope of this work).

This makes accordingly $2 \times 2N^2$ parameters to determine, the $2N^2$
VEV's $v, \hat v$ of the ${\mathfrak s}^0, {\mathfrak s}^3$'s
 and the $2N^2$ VEV's $\mu^3, \hat\mu^3$ of the neutral scalar composite
operators $<\bar q_i q_j>$.

\section{The Yukawa and kinetic Lagrangians}\label{section:kinyuk}
%================================================================

\subsection{Overview}
%---------------------

Yukawa couplings, originally devised to trigger fermion mass terms, are
built so as to be invariant by the (electro-)weak group.
They are not invariant  by the chiral group $U(2N)_L\times U(2N)_R$,
which also makes them  suitable to
trigger, through low energy theorems, the masses of $J=0$ scalar and
pseudoscalar mesons. The scalar potential 
is chosen to be $U(2N)_L \times U(2N)_R$ chirally invariant such that all
these states would be true Goldstones in the absence of Yukawa couplings
and only get ``soft'' masses in their presence, by the effect of chiral
symmetry breaking (since the weak group is a subgroup of the chiral group,
weak and chiral breaking are of course entangled). The only 
exceptions are the 3 Goldstones of the spontaneously broken
$SU(2)_L$, which should remain exactly massless and become the longitudinal
components of the 3 massive $W$'s. The scalar spectrum of the theory is
therefore composed of $8N^2-3$ pseudo-Goldstones bosons. Some are scalars,
including the ``Higgs'' boson and its avatars, the other are pseudoscalar
mesons, which  should fit those observed
experimentally. The latter should in particular reproduce well known 
symmetry patterns which, up to a good precision, fits them into
representations of a ``rotated'' flavor group (we call it rotated because
these bound states are made with mass eigenstates and not flavor
eigenstates).  As far as scalar mesons are concerned, 
no particular symmetry structure should be found, as observed
in their somewhat chaotic mass spectrum.

It may be opportune here to mention that ``low energy'' considerations,
like the PCAC (Partially Conserved Axial Current hypothesis) and 
Gell-Mann-Oakes-Renner (GMOR) relations  should be roughly trustable at
mass scales below a few GeV's, which is much smaller that the weak scale.
This is the case for 2 generations of quarks. However, when the top quark
comes into the game, they should be taken with great care. This is one of
the reasons why the realistic case of 3 generations is expected to be much
more cumbersome that the one dealt with in this note.

\subsection{The Yukawa Lagrangian}
%---------------------------------

\subsubsection{Its exact expression}
%-----------------------------------

Writing the most general such terms would mean coupling the two $SU(2)_L$ quark
doublets $\left(\begin{array}{c} u_L\cr d_L\end{array}\right)$ and
$\left(\begin{array}{c} c_L\cr s_L\end{array}\right)$ (and the 4
corresponding right-handed singlets) to the $2N^2 = 8$ available normalized
 $\Delta$
quadruplets (to generate masses for $d$-type quarks) and to their
corresponding $2N^2$ conjugate alter-ego's  $\;i\frac{T^2}{2}\Delta^\ast$
(to generate masses for the
$u$-type quarks).  This amounts to 64 couplings for 2 generations.

We drastically reduce their number down to 16 by comparison with what has been done in
the case of 1 generation \cite{Machet1}\cite{Machet2}. We  write them as an
``educated'' quadratic sum over the $N^2$ set of pairs of quadruplets made of one
 $\Delta_i$ and its parity-transformed $\hat\Delta_i$, $i=X,H,\Omega,\Xi$
\begin{equation}
{\cal L}_{Yuk}=
 \sum_{i=X,H,\Omega,\Xi} 
-\delta_i\, \Delta_i^\dagger\, [\Delta_i] -\delta_{i\hat i}\,
\Delta_i^\dagger\,
[\hat\Delta_i] -\kappa_{\hat i i}\, \hat\Delta_i^\dagger\, [\Delta_i]
-\hat\delta_i\,
\hat\Delta_i^\dagger\, [\hat\Delta_i].
\label{eq:Lyuk}
\end{equation}
In the formula (\ref{eq:Lyuk}), the $\Delta_i$'s and $\hat\Delta_i$'s
 stand for the complex $SU(2)_L$
doublets  of the type (\ref{eq:doublet}) expressed
 in terms of quarks bilinears that are built
 from the quadruplets displayed in
(\ref{eq:Xq}), (\ref{eq:Hq}), (\ref{eq:Omq}) and (\ref{eq:Xiq}), and
with their parity transformed.
The $[\Delta]_i$'s and $[\hat\Delta_i]$'s
are  the (same) doublets but expressed in terms of bosonic
fields with dimension $[mass]$
\begin{equation}
\begin{split}
 [X] = \left(\begin{array}{r} [X^1]-i[X^2]\cr
-([X^0]+[X^3])\end{array}\right)&,\quad
[H] = \left(\begin{array}{r} [H^1]-i[H^2]\cr
-([H^0]+[H^3])\end{array}\right),\cr
 [\Omega] = \left(\begin{array}{r} [\Omega^1]-i[\Omega^2]\cr
-([\Omega^0]+[\Omega^3])\end{array}\right)&,\quad
[\Xi] = \left(\begin{array}{r} [\Xi^1]-i[\Xi^2]\cr
-([\Xi^0]+[\Xi^3])\end{array}\right),
\end{split}
\end{equation}
and their parity transformed.

In the case of 1 generation, $i$ reduces to a single value and one recovers
the most general Yukawa couplings for $(u,d)$ quarks given in eqs.~(8) and
(9) of \cite{Machet2}, which is also the one of  the GSW model.
The expression (\ref{eq:Lyuk})  is its simplest generalization to 2
generations, in that it is
the sum of the 4 similar ``diagonal'' contributions corresponding to the the 4 pairs
$(\Delta_i, \hat\Delta_i), i=X,H,\Omega,\Xi$, discarding all
cross-couplings between different pairs $i\not=j$.

So written, ${\cal L}_{Yuk}$ couples the $2N$ quarks to all (pseudo-)scalar
fields in a very specific way.
Associated with the specific choice (\ref{eq:Xq}), (\ref{eq:Hq}),
(\ref{eq:Omq}) and (\ref{eq:Xiq}) for the quadruplets (any linear
combinations would a priori also be a suitable possibility), it has the 
property of maximally avoiding flavor changing neutral currents (FCNC's) at the
classical level.
Introducing a coupling like $H^\dagger [X]$ would indeed
generate at low energy  a 4-fermion coupling proportional to
$(\bar u\gamma^5 d)(\bar s\gamma^5 c)$ which carries unwanted
$u\to c$ and $d\to s$ transitions. 
The case of crossed $\Omega-\Xi$ couplings is less evident, apart from the
fact that it would generate classical transitions between $K^0+\bar K^0$
and $K^0-\bar K^0$. One can also argue that, formally, all quadruplets
being equivalent, there is no reason to cross-couple some of them and not
the others. We will show in this work and in the following ones that this choice
leads to consistent results.

\subsection{The kinetic Lagrangian for the scalar sector}
%--------------------------------------------------------

It is 
\begin{equation}
{\cal L}_{kin}=\sum_{i=X,H,\Omega,\Xi} D_\mu [\Delta_i]^\dagger D^\mu[\Delta_i]
+ D_\mu [\hat\Delta_i]^\dagger D^\mu[\hat\Delta_i],
\end{equation}
where $D_\mu$ is the covariant derivative with respect to the
(electro-)weak group.

The mass of the $W$'s is accordingly given by
\begin{equation}
m_W^2 =\frac{g^2}{4} \sum_{i=X,H,\Omega,\Xi}(v_i^2 + \hat v_i^2).
\end{equation}

\subsection{Choosing the quasi-standard Higgs doublet}\label{subsec:choice}
%--------------------------------------------------------------------------

We have to make a choice concerning which quadruplet contains the 3
true Goldstone bosons of the broken $SU(2)_L$.
If we choose a $({\mathfrak s},\vec {\mathfrak p})$ quadruplets,
2 charged and 1 neutral pseudoscalar mesons
will automatically disappear from the spectrum. This is disfavored since
all charged pseudoscalar mesons for 2 generations have been observed.
If we choose $\hat\Omega$ or $\hat\Xi$, $\hat\Omega^0$ or $\hat\Xi^0$ is
doomed to become the longitudinal $W^3_\parallel$; this is not good either
since these are interpolating fields for neutral kaons and $D$ mesons. We
have accordingly to decide between $\hat X$ and $\hat H$. Since it looks
better that the heaviest quark, the one that presumably enters into the
composition of the quasi-standard
Higgs boson, is called $c$ rather than $u$, we  choose
$\hat H$ as the ``quasi-standard'' Higgs doublet for 2 generations.

\section{Masses and orthogonality of charged pseudoscalar mesons. The
Cabibbo angle}
%================================================================

\subsection{The rise of mixing}
%------------------------------

By the nature of the quadruplets $\Omega, \hat\Omega, \Xi,\hat\Xi$, their 
``self-coupling'' occurring in the Yukawa Lagrangian triggers, through the
VEV's $v_\Omega, \hat v_\Omega, v_\Xi, \hat v_\Xi$, non-diagonal fermionic
mass terms $\bar u c, \bar c u, \bar d s, \bar s d$. It is then
straightforward to get an expression for the $\tan$ of twice the mixing
angles $\theta_u$ and $\theta_d$ in terms of Yukawa parameters.

This is however not our concern here and we shall only introduce the two
mixing angles $\theta_u$ and $\theta_d$ and  the quark mass
eigenstates $u_m, c_m, d_m, s_m$ as usual by ($c_u, s_u$ mean respectively
$\cos\theta_u$ and $\sin\theta_u$ {\em etc})
\begin{equation}
\left(\begin{array}{c}u \cr c\end{array}\right)=
\left(\begin{array}{rr} c_u & s_u \cr -s_u & c_u \end{array}\right)
\left(\begin{array}{c}u_m \cr c_m\end{array}\right),\quad
\left(\begin{array}{c}d \cr s\end{array}\right)=
\left(\begin{array}{rr} c_d & s_d \cr -s_d & c_d \end{array}\right)
\left(\begin{array}{c}d_m \cr s_m\end{array}\right).
\label{eq:mix1}
\end{equation}

We shall then work at the mesonic level by using low energy theorems.

\subsection{At low energy}
%-------------------------

The tools at our disposal are the statement that the divergences of axial
currents of massive quarks are suitable interpolating fields for the
corresponding mesons (PCAC) \cite{Dashen} and the Gell-Mann-Oakes-Renner relation
\cite{GMOR} which
evaluates 2-point functions of such divergences at low momentum
\footnote{See also \cite{Lee} and \cite{dAFFR} for general reviews.}.

They result, for example for the charged pions, into the 2 relations
\begin{equation}
\begin{split}
& i(m_u+m_d) \bar u_m\gamma^5 d_m = \sqrt{2} f_\pi m_\pi^2 \pi^+,\cr
& (m_u+m_d)<\bar u_m u_m + \bar d_m d_m>=2f_\pi^2 m_\pi^2,
\end{split}
\end{equation}
which evidently concern quark mass eigenstates.

With the help of these relations and equivalent, many entries of the
composite quadruplets can be expressed in terms of known ``particles'', in
particular charged pseudoscalar mesons $\pi^\pm, K^\pm, D^\pm, D_s^\pm$
\footnote{
For example, {\em if there was no mixing}, $X^+$ would write
$-i\frac{v_X}{f_\pi}\pi^+$. When mixing occurs, it becomes a linear
combination of $\pi^+, K^+, D^+, D_s^+$.}.
This leads to the bosonised forms of the kinetic terms and Yukawa Lagrangian,
valid at low energy for meson physics.

They are the ones that we use in the following and from which we request
the two conditions:\newline
$\ast$\  no crossed terms between different charged pseudoscalar mesons
should arise in the bosonised Yukawa Lagrangian;\newline
$\ast$\ the ratios of the  quadratic terms in the
Yukawa and kinetic Lagrangian for these  states provide their
$mass^2$.

We are careful to only use at this stage charged pseudoscalar mesons
because they are experimentally observed not to mix. This is not the case
for neutral pseudoscalars, the mixing pattern of which can be quite complex
(and should be predictable in principle in our approach).

\subsection{Notations}
%--------------------

Because this short note does not aim at determining all
parameters and because the solutions of the restricted set of equations that we shall
consider for our purpose  are mostly expressed in terms of the following
ones, we shall define, for each pair of VEV's
 $(\frac{v}{\sqrt{2}},\mu^3)$ or
$(\frac{\hat v}{\sqrt{2}}, \hat \mu^3)$, the ratio with dimension $[mass]^2$ 
\begin{equation}
\nu_i^2 = \frac {\sqrt{2}\,\mu_i^3}{v_i},\quad
\hat\nu_i^2 = \frac{\sqrt{2}\,\hat\mu_i^3}{\hat
v_i},\quad i=X,H,\Omega,\Xi.
\label{eq:nus}
\end{equation} 

A priori $<\bar u c> = <\bar c u>$ and $<\bar d s> = <\bar s d>$ such
that $\mu_\Xi^3=0$ and $\hat\mu_\Xi^3=0$. This does not mean however that
$v_\Xi$ or $\hat v_\Xi$ automatically vanishes.

We shall also use the following dimensionless parameters
\begin{equation}
b_i= \left(\frac{v_i}{\hat v_H}\right)^2,\quad \hat b_i = \left(\frac{\hat v_i}{\hat
v_H}\right)^2,\quad i=X, H, \Omega, \Xi,
\label{eq:bees}
\end{equation}
such that, by definition (in relation with our choice for the
``quasi-standard'' Higgs quadruplet that includes the 3 Goldstones of the
spontaneously broken $SU(2)_L$ symmetry, see subsection \ref{subsec:choice}) 
\begin{equation}
\hat b_H=1.
\label{eq:bh}
\end{equation}
We shall also use  the parameters
\begin{equation}
\frac{1}{\bar\nu_i^4} = \frac{1-b_i}{\nu_i^4}.
\label{eq:barnus}
\end{equation}

\subsection{Mesons quadratics : orthogonality}
%---------------------------------------------

\subsubsection{Starting conditions}
%----------------------------------

Charged pseudoscalar mesons only occur  in the ``non-hatted'' bosonised quadruplets
$X, H, \Omega, \Xi$.
The non-diagonal couplings between them in the bosonised Yukawa Lagrangian
are proportional to the following expressions that should accordingly vanish
($c_{u-d}$ stands for $\cos(\theta_u-\theta_d)$ {\em etc})
\begin{equation}
\begin{split}
(\pi-K)&: 
\delta_X \frac{c_uc_dc_us_d}{\nu_X^4} - \delta_H \frac{s_us_ds_uc_d}{\nu_H^4} 
-\frac{\delta_\Omega}{2} \frac{s_{u+d}c_{u+d}}{\nu_\Omega^4} 
+\frac{\delta_\Xi}{2} \frac{s_{u-d}c_{u-d}}{\nu_\Xi^4} =0,\cr
(\pi-D)&:
\delta_X \frac{c_uc_ds_uc_d}{\nu_X^4} -\delta_H \frac{s_us_dc_us_d}{\nu_H^4} 
-\frac{\delta_\Omega}{2} \frac{s_{u+d}c_{u+d}}{\nu_\Omega^4} -
\frac{\delta_\Xi}{2} \frac{s_{u-d}c_{u-d}}{\nu_\Xi^4}
=0,\cr
(\pi-D_s)&: 
\delta_X \frac{s_us_dc_uc_d}{\nu_X^4} +\delta_H \frac{s_us_dc_uc_d}{\nu_H^4} 
-\frac{\delta_\Omega }{2} \frac{s_{u+d}^2}{\nu_\Omega^4} +
\frac{\delta_\Xi}{2} \frac{s_{u-d}^2}{\nu_\Xi^4}
=0,\cr
(K-D)&:
\delta_X \frac{c_us_ds_uc_d}{\nu_X^4} +\delta_H
\frac{s_uc_dc_us_d}{\nu_H^4} 
+\frac{\delta_\Omega}{2} \frac{c_{u+d}^2}{\nu_\Omega^4} -
\frac{\delta_\Xi}{2} \frac{c_{u-d}^2}{\nu_\Xi^4}
=0,\cr
(K-D_s)&:
\delta_X \frac{c_us_ds_us_d}{\nu_X^4} -\delta_H \frac{s_uc_dc_uc_d}{\nu_H^4} 
+\frac{\delta_\Omega }{2} \frac{s_{u+d}c_{u+d}}{\nu_\Omega^4} +
\frac{\delta_\Xi}{2} \frac{s_{u-d}c_{u-d}}{\nu_\Xi^4}
=0,\cr
(D-D_s)&:
\delta_X \frac{s_uc_ds_us_d}{\nu_X^4} -\delta_H \frac{c_us_dc_uc_d}{\nu_H^4}
+\frac{\delta_\Omega}{2} \frac{s_{u+d}c_{u+d}}{\nu_\Omega^4} -
\frac{\delta_\Xi}{2} \frac{s_{u-d}c_{u-d}}{\nu_\Xi^4}
=0.
\label{eq:ndiag0}
\end{split}
\end{equation}

\subsection{Basics for the scalar potential. Connecting the
$\boldsymbol{\delta_i}$'s.}
%------------------------------------------------

Relations between $\delta_X, \delta_H, \delta_\Omega, \delta_\Xi$ can be
obtained by minimizing the effective potential $V_{eff}(\Delta_i)$ obtained
by subtracting the bosonised Yukawa Lagrangian
\footnote{This symmetric and hermitian form is obtained by simply rewriting
all terms in the Yukawa Lagrangian (\ref{eq:Lyuk}) in terms of fields of
dimension $mass$ like in \cite{Machet2}.}
\begin{equation}
{\cal L}_{Yuk}^{bos}=
 \sum_{i=X,H,\Omega,\Xi} 
-\delta_i\, [\Delta_i]^\dagger\, [\Delta_i]
-\frac12(\delta_{i\hat i}+\kappa_{\hat i i})\,
\left([\Delta_i]^\dagger\, [\hat\Delta_i] + [\hat\Delta_i]^\dagger\, [\Delta_i]
\right)
-\hat\delta_i\, [\hat\Delta_i]^\dagger\, [\hat\Delta_i]
\label{eq:Lyukbos}
\end{equation}
 to the scalar potential $V(\Delta_i)$ suitably chosen.
To this purpose, it is most efficient to work in ``flavor space'', which
means here using the components $\Delta_i^0, \Delta_i^3, \Delta_i^+,
\Delta_i^-$ of
each quadruplet $\Delta_i$ and not the meson fields like $\pi, K \ldots$ 

There again, the choice of $V$ is important. The most general scalar
potential for $2N^2=8$ Higgs multiplets has a large number of parameters.
However, as we already mentioned, we choose it to be $U(2N)_L \times
U(2N)_R$ chirally invariant and such that no nonphysical transition between
known particles, nor any unrealistic mass splitting gets induced at the
classical level. These requirements lead to an extremely simple form, like
for the Yukawa Lagrangian, which is
\begin{equation}
V= -\frac{m_H^2}{2} \sum_i\Delta_i^\dagger \Delta_i + \frac{\lambda_H}{4}
\sum_i(\Delta_i^\dagger \Delta_i)^2,\quad i=X,H,\Omega,\Xi, \hat X,\hat H, \hat\Omega,
\hat\Xi.
\label{eq:V}
\end{equation}
It only involves 2 parameters, $m_H^2$ and $\lambda_H$.
The effective potential $V_{eff}=V-{\cal L}_{Yuk}$ therefore involves
18 unknown parameters.

The bosonised Yukawa Lagrangian gets simplified by requesting that charged
pseudoscalar and scalar mesons do not couple at the classical level. This
requires
\begin{equation}
\delta_{i\hat i} + \kappa_{\hat i i}=0,\quad i=X, H, \Omega, \Xi.
\end{equation} 
Minimizing $V_{eff}$ at the values $<X^0>=\frac{v_X}{\sqrt{2}},
<\hat H^3>=\frac{\hat v_H}{\sqrt{2}} \ldots$ yields then $2\times 4 = 8$
 equations of the type
\begin{equation}
m_H^2 = \lambda_H \frac{v_i^2}{2} + 2\delta_i,\ldots\quad
m_H^2 = \lambda_H\frac{\hat v_i^2}{2} + 2\hat\delta_i,\ldots
\end{equation}
One among them is special, the one related to the ``quasi-standard'' Higgs
doublet $\hat H$. That the 3 Goldstone bosons of the broken chiral symmetry
that it contains stay as the 3 true Goldstones of the spontaneously broken
$SU(2)_L$ requires in particular
\begin{equation}
\hat\delta_H=0,
\end{equation}
which entails
\begin{equation}
m_H^2 = \lambda_H \frac{\hat v_H^2}{2},
\end{equation}
and thus
\begin{equation}
\lambda_H= \frac{4\delta_i}{\hat v_H^2-v_i^2}=\frac{4\hat\delta_i}{\hat
v_H^2 -\hat v_i^2}.
\end{equation}
Let us  define $\delta$ such that
\footnote{The mass scale set by $\delta$, tightly connected with the mass
of the ``quasi-standard'' Higgs boson, can be evaluated by looking at
neutral kaons and $D$ mesons. We do not need it here and therefore delay
its presentation to \cite{Machet4}.}
\begin{equation}
\lambda_H=\frac{4\delta}{\hat v_H^2} \Rightarrow m_H^2=2\delta.
\end{equation}
Then
\begin{equation}
\delta_i =\delta(1-b_i),\quad \hat\delta=\delta(1-\hat b_i).
\label{eq:deltarel}
\end{equation}

\subsection{Solution of the equations (\ref{eq:ndiag0})}
%-------------------------------------------------------
Using the relations (\ref{eq:deltarel}) between the $\delta_i$
 and (\ref{eq:barnus}), $\delta \not=0$ can be factored out and equations
(\ref{eq:ndiag0})  rewrite
\begin{equation}
\begin{split}
& (a) : 
\frac{c_uc_dc_us_d}{\bar\nu_X^4} -\frac{s_us_ds_uc_d}{\bar\nu_H^4} 
-\frac12 \frac{s_{u+d}\,c_{u+d}}{\bar\nu_\Omega^4} +\frac12
\frac{s_{u-d}\,c_{u-d}}{\bar\nu_\Xi^4}
=0,\cr
& (b) :
\frac{c_uc_ds_uc_d}{\bar\nu_X^4} -\frac{s_us_dc_us_d}{\bar\nu_H^4} 
-\frac12 \frac{s_{u+d}\,c_{u+d}}{\bar\nu_\Omega^4} -\frac12
\frac{s_{u-d}\,c_{u-d}}{\bar\nu_\Xi^4}
=0,\cr
&  (c) :
\frac{s_us_dc_uc_d}{\bar\nu_X^4} +\frac{s_us_dc_uc_d}{\bar\nu_H^4} 
-\frac12 \frac{s_{u+d}^2}{\bar\nu_\Omega^4} +\frac12
\frac{s_{u-d}^2}{\bar\nu_\Xi^4}
=0,\cr
& (d) :
\frac{c_us_ds_uc_d}{\bar\nu_X^4} +\frac{s_uc_dc_us_d}{\bar\nu_H^4} 
+\frac12 \frac{c_{u+d}^2}{\bar\nu_\Omega^4} -\frac12
\frac{c_{u-d}^2}{\bar\nu_\Xi^4}
=0,\cr
& (e) :
\frac{c_us_ds_us_d}{\bar\nu_X^4} -\frac{s_uc_dc_uc_d}{\bar\nu_H^4} 
+\frac12 \frac{s_{u+d}\,c_{u+d}}{\bar\nu_\Omega^4} +\frac12
\frac{s_{u-d}\,c_{u-d}}{\bar\nu_\Xi^4}
=0,\cr
& (f) :
\frac{s_uc_ds_us_d}{\bar\nu_X^4} -\frac{c_us_dc_uc_d}{\bar\nu_H^4} 
+\frac12 \frac{s_{u+d}\,c_{u+d}}{\bar\nu_\Omega^4} -\frac12
\frac{s_{u-d}\,c_{u-d}}{\bar\nu_\Xi^4}
=0,
\label{eq:ndiag1}
\end{split}
\end{equation}
or, equivalently, by recombining the equation
\begin{equation}
\begin{split}
& (a)+(f) : s_{2d}\left(\frac{1}{\bar\nu_X^4}-\frac{1}{\bar\nu_H^4}\right)=0,\cr
& (a)-(f) :
s_{2d}c_{2u}\left(\frac{1}{\bar\nu_X^4}+\frac{1}{\bar\nu_H^4}\right)-\frac{s_{2(u+d)}}{\bar\nu_\Omega^4}+\frac{s_{2(u-d)}}{\bar\nu_\Xi^4}=0,\cr
& (b)-(e) :
s_{2u}c_{2d}\left(\frac{1}{\bar\nu_X^4}+\frac{1}{\bar\nu_H^4}\right)-\frac{s_{2(u+d)}}{\bar\nu_\Omega^4}-\frac{s_{2(u-d)}}{\bar\nu_\Xi^4}=0,\cr
& (b)+(e) : s_{2u}\left(\frac{1}{\bar\nu_X^4} -\frac{1}{\bar\nu_H^4}\right)=0,\cr
& (c)-(d) : \frac{1}{\bar\nu_\Omega^4}-\frac{1}{\bar\nu_\Xi^4}=0,\cr
& (c)+(d) : s_{2u}s_{2d}\left(\frac{1}{\bar\nu_X^4} +
\frac{1}{\bar\nu_H^4}\right)+\frac{c_{2(u+d)} }{\bar\nu_\Omega^4}
-\frac{c_{2(u-d)}}{\bar\nu_\Xi^4}=0.
\label{eq:ndiag+3}
\end{split}
\end{equation}
The solution of (\ref{eq:ndiag+3}) is 
\begin{equation}
\frac{1}{\bar\nu_X^4} = \frac{1}{\bar\nu_H^4} =
\frac{1}{\bar\nu_\Omega^4}=\frac{1}{\bar\nu_\Xi^4}
\stackrel{(\ref{eq:barnus})}{\Leftrightarrow}
\frac{1-b_X}{\nu_X^4}=\frac{1-b_H}{\nu_H^4}
=\frac{1-b_\Omega}{\nu_\Omega^4}=\frac{1-b_\Xi}{\nu_\Xi^4}.
\label{eq:solnu1}
\end{equation}

\subsection{Mesons quadratics : masses}
%-------------------------------------

From the ratios of the terms quadratic in the meson fields in the effective
potential and in the kinetic terms, using (\ref{eq:deltarel}) and
(\ref{eq:bees}) one gets
\begin{equation}
\begin{split}
 m_{\pi^\pm}^2 &=\delta \frac{ \left(1-b_X\right)
(\frac{c_uc_d}{\nu_X^2})^2 +
 \left(1-b_H\right)(\frac{s_us_d}{\nu_H^2})^2 +
\left(1-b_\Omega\right) \frac12(\frac{s_{u+d}}{\nu_\Omega^2})^2
 + \left(1-b_\Xi\right)\frac12(\frac{s_{u-d}}{\nu_\Xi^2})^2 }
{(\frac{c_uc_d}{\nu_X^2})^2 + (\frac{s_us_d}{\nu_H^2})^2 +
\frac12(\frac{s_{u+d}}{\nu_\Omega^2})^2 + \frac12(\frac{s_{u-d}}{\nu_\Xi^2})^2 },\cr
 m_{K^\pm}^2 &= \delta \frac{ \left(1-b_X\right)
(\frac{c_us_d}{\nu_X^2})^2 +
 \left(1-b_H\right)(\frac{s_uc_d}{ \nu_H^2})^2 +
\left(1-b_\Omega\right) \frac12(\frac{c_{u+d}}{\nu_\Omega^2})^2
+ \left(1-b_\Xi\right)\frac12(\frac{c_{u-d}}{\nu_\Xi^2})^2 }
{(\frac{c_us_d}{\nu_X^2})^2 + (\frac{s_uc_d}{ \nu_H^2})^2 +
\frac12(\frac{c_{u+d}}{\nu_\Omega^2})^2 +
\frac12(\frac{c_{u-d}}{\nu_\Xi^2})^2},\cr
 m_{D^\pm}^2 &= \delta \frac{ \left(1-b_X\right)
(\frac{s_uc_d}{\nu_X^2})^2 +
 \left(1-b_H\right)(\frac{c_us_d}{ \nu_H^2})^2 +
\left(1-b_\Omega\right) \frac12(\frac{c_{u+d}}{\nu_\Omega^2})^2
+ \left(1-b_\Xi\right)\frac12(\frac{c_{u-d}}{\nu_\Xi^2})^2 }
{(\frac{s_uc_d}{\nu_X^2})^2 + (\frac{c_us_d}{ \nu_H^2})^2 +
\frac12(\frac{c_{u+d}}{\nu_\Omega^2})^2 +
\frac12(\frac{c_{u-d}}{\nu_\Xi^2})^2 },\cr
 m_{D_s^\pm}^2 &= \delta \frac{ \left(1-b_X\right)
(\frac{s_us_d}{\nu_X^2})^2 +
 \left(1-b_H\right)(\frac{c_uc_d}{ \nu_H^2})^2 +
\left(1-b_\Omega\right) \frac12(\frac{s_{u+d}}{\nu_\Omega^2})^2
+ \left(1-b_\Xi\right)\frac12(\frac{s_{u-d}}{\nu_\Xi^2})^2 }
{(\frac{s_us_d}{\nu_X^2})^2 + (\frac{c_uc_d}{ \nu_H^2})^2 +
\frac12(\frac{s_{u+d}}{\nu_\Omega^2})^2 +
\frac12(\frac{s_{u-d}}{\nu_\Xi^2})^2 },
\label{eq:diag+0}
\end{split}
\end{equation}
which rewrites, using (\ref{eq:solnu1})
\begin{equation}
\begin{split}
& m_{\pi^\pm}^2 =\frac{\delta/\bar\nu_X^4}{(c_uc_d/\nu_X^2)^2 + (s_us_d/ \nu_H^2)^2 +
\frac12 (s_{u+d}/\nu_\Omega^2)^2 + \frac12 (s_{u-d}/\nu_\Xi^2)^2},\cr
& m_{K^\pm}^2 = \frac{\delta/\bar\nu_X^4}{(c_us_d/\nu_X^2)^2 + (s_uc_d/ \nu_H^2)^2 +
\frac12 (c_{u+d}/\nu_\Omega^2)^2 +
\frac12 (c_{u-d}/\nu_\Xi^2)^2},\cr
& m_{D^\pm}^2 =  \frac{\delta/\bar\nu_X^4}{(s_uc_d/\nu_X^2)^2 + (c_us_d/ \nu_H^2)^2 +
\frac12 (c_{u+d}/\nu_\Omega^2)^2 +
\frac12 (c_{u-d}/\nu_\Xi^2)^2},\cr
& m_{D_s^\pm}^2 =  \frac{\delta/\bar\nu_X^4}{(s_us_d/\nu_X^2)^2 + (c_uc_d/
\nu_H^2)^2 + \frac12 (s_{u+d}/\nu_\Omega^2)^2 +
\frac12 (s_{u-d}/\nu_\Xi^2)^2}.
\label{eq:diag+1}
\end{split}
\end{equation}
Eqs.~(\ref{eq:diag+1}) entail
\begin{equation}
\begin{split}
& \delta \left( +\frac{1}{m_{\pi^\pm}^2}+\frac{1}{m_{K^\pm}^2} +
\frac{1}{m_{D^\pm}^2}+\frac{1}{m_{D_s^\pm}^2}\right)
=\frac{1}{1-b_X} +\frac{1}{1-b_H}+\frac{1}{1-b_\Omega}+ \frac{1}{1-b_\Xi},\cr
& \delta \left( +\frac{1}{m_{\pi^\pm}^2}-\frac{1}{m_{K^\pm}^2} +
\frac{1}{m_{D^\pm}^2}-\frac{1}{m_{D_s^\pm}^2}\right)
=c_{2d}\left(\frac{1}{1-b_X} - \frac{1}{1-b_H} \right),\cr
& \delta \left( +\frac{1}{m_{\pi^\pm}^2}+\frac{1}{m_{K^\pm}^2} -
\frac{1}{m_{D^\pm}^2}-\frac{1}{m_{D_s^\pm}^2}\right)
=c_{2u}\left(\frac{1}{1-b_X} - \frac{1}{1-b_H} \right),\cr
& \delta \left( +\frac{1}{m_{\pi^\pm}^2}-\frac{1}{m_{K^\pm}^2} -
\frac{1}{m_{D^\pm}^2}+\frac{1}{m_{D_s^\pm}^2}\right)
= c_{2u}c_{2d}\left(\frac{1}{1-b_X} + \frac{1}{1-b_H} \right)
-\frac{c_{2(u+d)}}{1-b_\Omega} -\frac{c_{2(u-d)}}{1-b_\Xi}.
\end{split}
\label{eq:diag+2}
\end{equation}

\subsection{The Cabibbo angle}
%-----------------------------
From the second and third equations of (\ref{eq:diag+2}) one gets,
independently of the scale $\delta$
\begin{equation}
\frac{c_{2u}-c_{2d}}{c_{2u}+c_{2d}}
\equiv\tan(\theta_d+\theta_u)\tan(\theta_d-\theta_u)
=\frac
{\frac{1}{m_{K^\pm}^2}-\frac{1}{m_{D^\pm}^2}}
{\frac{1}{m_{\pi^\pm}^2}-\frac{1}{m_{D_s^\pm}^2}},
\label{eq:cab1}
\end{equation}
which vanishes either at the chiral limit $m_\pi\to 0$ or when $m_K=m_D$.

By the freedom to make an arbitrary flavor rotation on $(u,c)$ quarks,
one can align flavor and mass eigenstates in this sector and, therefore,
tune $\theta_u \to 0$.  $\theta_d$ becomes then the
Cabibbo angle $\theta_c$ which is given by
\begin{equation}
\boxed{
\tan^2\theta_c=
\displaystyle\frac{\displaystyle\frac{1}{m_{K^\pm}^2}-\displaystyle\frac{1}{m_{D^\pm}^2}}
{\displaystyle\frac{1}{m_{\pi^\pm}^2}-\displaystyle\frac{1}{m_{D_s^\pm}^2}}
\approx\frac{m_{\pi^\pm}^2}{m_{K^\pm}^2}\left(1-\frac{m_{K^\pm}^2}{m_{D^\pm}^2}
+\frac{m_{\pi^\pm}^2}{m_{D_s^\pm}^2}\right)
}\qquad q.e.d.
\label{eq:tantheta}
\end{equation}
Numerically, it corresponds to $\theta_c
\approx .27$, to be compared with the measured $\approx .23$.

\section{Conclusion and prospects}
%=================================

With the example of the Cabibbo angle, we have shown that the extension
that we propose for the GSW model allows calculations that have long been
sought for
\footnote{This long quest started with ref.\cite{Weinberg}. Since then a
large literature has been devoted to it, looking mainly for connections
between quark masses and mixing angles. An extensive quotation lies beyond
the scope of this note.}.  This angle we determined from the
sole physical data concerning the masses and orthogonality of the 4 types of
charged pseudoscalar mesons $\pi^\pm, K^\pm, D^\pm$ and $D_s^\pm$, such
that we had only to exploit a small part of the physical information
available concerning pseudoscalar mesons.

While the Higgs sector of the GSW model has been maximally extended by
including in it all $J=0$ mesons expected for a given number of generations
of quarks, the Yukawa Lagrangian and the scalar potential have been reduced to
very simple expressions by requirements of invariance and to avoid classical
unwanted phenomena like FCNC, non-existing crossed couplings between known
states and unrealistic mass differences (like, in the case of 1 generation,
$\pi^+-\pi^0$ mass difference which originates neither from $m_d \not= m_u$
nor from electromagnetic corrections). This makes this extension the
simplest, minimal and most natural one, showing that these criteria may be at
work once more in nature.

Obtaining a sensible expression for the Cabibbo angle suggests
that this direction is worth  detailed investigations.
I shall present in a forthcoming work \cite{Machet4}, still for 2
generations, the values of all VEV's, the masses of all
Higgs bosons and  their couplings to gauge bosons and to fermions.

\bigskip
\underline{\em Acknowledgments}: {\em a special thank is due to P. Slavich whose
expert eye immediately detected an erroneous $1/2$ in a draft of
this work.}

{\baselineskip 10pt
\vskip 10mm
\centerline{$\ast$}
\centerline{$\ast$\quad$\ast$}
}

%%%%%%%%%%%%%%%%%%%%%%%%%%%%%%%%%%%%%%%%%%%%%%%%%%%%%%%%%%%%%%%%%%%%%%%%%%%%%%
%\newpage

%\vfill

\begin{em}

\end{em}

{\baselineskip 10pt
\vskip 10mm
\centerline{$\ast$}
\centerline{$\ast$\quad$\ast$}
}
\end{document}